\newcommand*{\ATLASLATEXPATH}{}
\documentclass[cernpreprint, atlasdraft=true, texlive=2016, UKenglish]{\ATLASLATEXPATH atlasdoc}
 
\usepackage[subfigure]{\ATLASLATEXPATH atlaspackage}
\usepackage{longtable}
\usepackage{multicol}
\usepackage{tabularx}
\usepackage{adjustbox}
\usepackage{dcolumn}
\usepackage{colortbl}
\usepackage{listings}
\usepackage{floatrow}
\usepackage{\ATLASLATEXPATH atlasbiblatex}
 
\usepackage{\ATLASLATEXPATH atlasphysics}
 
\graphicspath{{logos/}{figures/}}
 
\usepackage{ANA-TOPQ-2020-02-PAPER-defs}

\AtlasDOI{10.1016/j.physletb.2020.135797}
\AtlasJournalRef{Phys. Lett. B 810 (2020) 135797}
\AtlasTitle{Measurement of the $t\bar{t}$ production cross-section in the lepton+jets channel at $\sqrt{s}=13\;$TeV with the ATLAS experiment }

\AtlasAbstract{
The top anti-top quark production cross-section is measured in the lepton+jets channel using proton--proton collision data at a centre-of-mass energy of $\sqrt{s}=13$~\tev~collected with the ATLAS detector at the LHC.
The dataset corresponds to an integrated luminosity of 139~fb$^{-1}$.
Events with exactly one charged lepton and four or more jets in the final state,
with at least one jet containing $b$-hadrons, are used to determine the $t\bar{t}$ production
cross-section through a profile-likelihood fit.
The inclusive cross-section is measured to be
${\sigma_{\text{inc}} =
830 \pm 0.4~\text{(stat.)}\pm 36~\text{(syst.)}\pm 14~\text{(lumi.)}~\mathrm{pb}}$
with a relative uncertainty of 4.6\%.
The result is consistent with theoretical calculations at next-to-next-to-leading
order in perturbative QCD.
The fiducial $t\bar{t}$ cross-section within the experimental acceptance is also measured.
}
 
\author{The ATLAS Collaboration}
 
\AtlasRefCode{TOPQ-2020-02}
 
\PreprintIdNumber{CERN-EP-2020-096}
 
\AtlasDate{\today}
 
\journal{PLB}
\hypersetup{pdftitle={ATLAS document},pdfauthor={The ATLAS Collaboration}}
 
\begin{document}
 
\maketitle
 
\section{Introduction}
\label{sec:intro}
 
The top quark is the heaviest elementary particle in the Standard Model (SM), with a
mass $m_t$ close to the electroweak symmetry breaking scale~\cite{TOPQ-2017-03, CMS-TOP-14-022}.
Studies of top-quark production and decays provide a precise
probe of the SM as well as its extensions~\cite{Buckley:2015eft}.
At the CERN Large Hadron Collider (LHC), top quarks are primarily produced in quark--antiquark
pairs (\ttbar) and form an important background in many searches for physics
beyond the SM.
Thus, a precise measurement of the \ttbar cross-section, and comparison with
theoretical predictions of high precision, are a critical part of the LHC physics programme.
 
A theoretical calculation of the \ttbar cross-section, $\sigma_{\ttbar}$, is available at next-to-next-to-leading
order (NNLO) in quantum chromodynamics (QCD). It includes the resummation of the
next-to-next-to-leading logarithmic (NNLL)
soft-gluon terms~\cite{Cacciari:2011hy,Baernreuther:2012ws,Czakon:2012zr,Czakon:2012pz,
Czakon:2013goa,Catani:2019iny} and predicts
\(\sigma_{\ttbar} = 832 _{-29}^{+20}~\text{(scale)} \pm 35~(\text{PDF} + \alphas)~\mathrm{pb} \)
in proton--proton ($pp$) collisions at a centre-of-mass energy of 13~\TeV, as calculated by
the \texttt{Top++} (v2.0) program~\cite{Czakon:2011xx}, using the MSTW2008 NNLO PDF set~\cite{Martin:2009iq, Martin:2009bu} as the central PDF set and assuming $m_t=172.5$~\GeV.
The scale uncertainty was determined from the envelope of predictions with the QCD
renormalisation and factorisation scales varied independently up or down by a factor of two.
The combined uncertainty due to the parton distribution functions (PDFs) and the strong coupling constant, $\alphas$,
was calculated following the PDF4LHC prescription~\cite{Botje:2011sn} with the
MSTW2008 NNLO, CT10 NNLO~\cite{Lai:2010vv,Gao:2013xoa} and
NNPDF2.3 5fFFN NNLO~\cite{Ball:2012cx} PDF sets.
 
Measurements of inclusive $\sigma_{\ttbar}$ at \num{7}, \num{8} and \SI{13}{\TeV} were performed by both the
ATLAS~\cite{TOPQ-2013-04, TOPQ-2015-18, Aad:2019hzw} and CMS~\cite{CMS-TOP-13-004, CMS-TOP-12-006,
CMS-TOP-16-005, Sirunyan:2017uhy, Sirunyan:2018goh} collaborations. All measurements are consistent with NNLO+NNLL QCD predictions.
Additionally, the CMS Collaboration performed a measurement of $\sigma_{\ttbar}$
at $\sqrt{s}=5.02~\tev$~\cite{Sirunyan:2017ule}.
At $\sqrt{s} = \SI{13}{\TeV}$, the ATLAS Collaboration
used a data sample
of 36.1~fb$^{-1}$ and events with an opposite-charge electron--muon pair in the final state to obtain
$\sigma_{t\bar{t}} = 826.4 \pm 3.6~\text{(stat.)}~\pm 11.5~\text{(syst.)}~\pm 15.7~\text{(lumi.)}~\pm 1.9~\text{(beam)}$ pb~\cite{TOPQ-2018-17}, giving a total relative uncertainty of 2.4\%.
 
This Letter documents measurements of the \ttbar cross-sections in the
full phase space (inclusive) and in a phase space defined to be close to the experimental measurement range (fiducial)
at $\sqrt{s} = 13~\tev$, using the full $pp$ dataset collected during 2015--2018.
It targets the lepton+jets \ttbar decay mode, where one $W$~boson originating
from the top quark decays leptonically and the other $W$~boson decays hadronically,
i.e.~\(\ttbar \ \rightarrow W^+ W^- b\bar{b} \rightarrow \ell \nu q\bar{q}' b\bar{b}\),
producing a final state
with one high-momentum electron or muon and four jets, two of which
are $b$-quark-initiated jets.\footnote{Events involving $W \rightarrow \tau\nu$ decays with a
subsequent decay of the $\tau$-lepton into $e\nu_e\nu_{\tau}$ or $\mu\nu_{\mu}\nu_{\tau}$
are included in the signal.} A small contribution from \ttbar events with both $W$~bosons decaying
leptonically producing the same final state due to one lepton being out of acceptance
is treated as signal.
A profile-likelihood fit to data in three non-overlapping regions is employed
to perform the measurement.
 
The study presented in this letter probes a final state that is complementary to
the one explored in Ref.~\cite{TOPQ-2018-17} and is sensitive to different \ttbar
modelling uncertainties, e.g. uncertainties related to quark jets,
the understanding of which is mandatory for a large number of top-quark
precision measurements and searches beyond the SM.
 
 
\section{ATLAS detector}
\label{sec:detector}
 
\newcommand{\AtlasCoordFootnote}{ATLAS uses a right-handed coordinate system with its origin at the nominal interaction point (IP) in the centre of the detector and the \(z\)-axis along the beam pipe.
The \(x\)-axis points from the IP to the centre of the LHC ring, and the \(y\)-axis points upwards.
Cylindrical coordinates \((r,\phi)\) are used in the transverse plane, \(\phi\) being the azimuthal angle around the \(z\)-axis.
The pseudorapidity is defined in terms of the polar angle \(\theta\) as \(\eta = -\ln \tan(\theta/2)\).
Angular distance is measured in units of \(\Delta R \equiv \sqrt{(\Delta\eta)^{2} + (\Delta\phi)^{2}}\).}
ATLAS~\cite{PERF-2007-01,Abbott:2018ikt,CERN-LHCC-2012-009} is a multipurpose particle detector designed with a forward--backward symmetric cylindrical geometry and nearly full \(4\pi\) coverage in solid angle.\footnote{\AtlasCoordFootnote}
It consists of an inner tracking detector surrounded by a thin superconducting solenoid providing a \SI{2}{\tesla} axial magnetic field, electromagnetic and hadronic calorimeters, and a muon spectrometer.
The inner tracking detector covers the pseudorapidity range \(|\eta| < 2.5\) and is composed of silicon pixel, silicon microstrip, and transition radiation tracking (TRT) detectors.
Lead/liquid-argon (LAr) sampling calorimeters provide electromagnetic (EM) energy measurements with high granularity.
Hadronic calorimetry is provided by the steel/scintillator-tile calorimeter covering the central pseudorapidity range (\(|\eta| < 1.7\)).
The endcap and forward regions are instrumented with LAr calorimeters for both the EM and hadronic energy measurements up to \(|\eta| = 4.9\).
The muon spectrometer surrounds the calorimeters and is based on three large air-core toroidal superconducting magnets with eight coils each.
The field integral of the toroids ranges between \num{2.0} and \SI{6.0}{\tesla\metre} across most of the detector.
The muon spectrometer includes a system of precision tracking chambers and fast detectors for triggering.
A two-level trigger system is used to select events.
The first-level trigger is implemented in hardware and uses a subset of the detector information
to keep the accepted event rate below \SI{100}{\kHz}~\cite{Aaboud:2016leb}.
This is followed by a software-based trigger that reduces the accepted event rate to \SI{1}{\kHz} on average.
 
\section{Data and simulation samples}
\label{sec:dataMC}
 
The analysis is performed using the full Run 2 LHC $pp$ collision data sample
at \( \sqrt{s} = 13~\si{\TeV}\) recorded by the ATLAS detector,
corresponding to an integrated luminosity of \(139\textrm{ fb}^{-1}\)
after data quality requirements~\cite{Aad:2019hmz} are imposed.
Events are required to pass a single-electron or single-muon trigger
with thresholds that were progressively raised during the data collection period
to account for the increase of instantaneous luminosity.
 
Monte Carlo (MC) simulations are used to optimise the analysis and to
evaluate acceptances, efficiencies and uncertainties in \ttbar signal and
all backgrounds except for the multijet background that is estimated using
a data-driven technique.
The effect of multiple interactions in the same and neighbouring bunch crossings (pile-up) was modelled by overlaying the original hard-scattering event with simulated inelastic $pp$
events generated by \textsc{Pythia 8.186}~\cite{Sjostrand:2007gs} using the NNPDF2.3 LO set of PDFs~\cite{Ball:2012cx} and parameter values set according to the A3 tune~\cite{ATL-PHYS-PUB-2016-017}.
 
The production of \ttbar\ events was modelled using the
next-to-leading-order (NLO) matrix element (ME) implemented in the
HVQ program~\cite{Frixione:2007nw,Frixione:2007nu} from the \POWHEGBOX~v2~\cite{Nason:2004rx,Frixione:2007vw,Alioli:2010xd} generator
with the NNPDF3.0 NLO~\cite{Ball:2014uwa} PDF and the $h_\text{damp}$ parameter
set to 1.5~$m_t$~\cite{ATL-PHYS-PUB-2016-020}.\footnote{The $h_\text{damp}$ parameter controls the transverse momentum, \pt,
of the first additional emission beyond the leading-order Feynman diagram in the parton shower and
therefore regulates the high-\pt\ emission against which the \ttbar\ system recoils.}
The \ttbar sample is normalised to the NNLO+NNLL cross-section.
The single-top-quark $t$-channel, $s$-channel and $tW$ associated production processes were also modelled at NLO
in QCD using \POWHEGBOX~v2. For all top-quark processes, \textsc{Pythia 8.230}~\cite{Sjostrand:2014zea}, using
the A14 tune~\cite{ATL-PHYS-PUB-2014-021} and the NNPDF2.3 LO PDF
set, was interfaced to \POWHEGBOX~v2 to simulate the parton
shower and hadronisation.
The diagram removal scheme~\cite{Frixione:2008yi} was employed in the $tW$ simulation
to handle the interference with \ttbar\ production~\cite{ATL-PHYS-PUB-2016-020}.
 
The $V$+jets ($V=W,Z$) backgrounds were simulated with
the \sherpa~v2.2.1~\cite{Bothmann:2019yzt} generator using NLO-accurate MEs
for up to two jets, and MEs accurate to leading order (LO)
for up to four jets calculated with the Comix~\cite{Gleisberg:2008fv}
and OpenLoops~\cite{Cascioli:2011va,Denner:2016kdg} libraries. They
were matched with the \sherpa parton shower~\cite{Schumann:2007mg} using the MEPS@NLO
prescription~\cite{Hoeche:2011fd,Hoeche:2012yf,Catani:2001cc,Hoeche:2009rj}
and the tune developed by the \sherpa authors.
Diboson production was generated using \sherpa~v2.2.2 with
MEs computed at NLO accuracy in QCD for up to one additional parton
and at LO accuracy for up to three additional partons.
The NNPDF3.0 NNLO PDF set~\cite{Ball:2014uwa} was used for the $V$+jets and diboson samples.
The productions of $\ttbar H$ and $\ttbar V$ events were modelled at NLO using the \POWHEGBOX~v2
and \MGMCatNLO~v2.3.3~\cite{Alwall:2014hca} generators, respectively, with
the NNPDF3.0 NLO PDF set.
\textsc{Pythia 8.230} with the A14 tune and the NNPDF2.3 LO PDF was used to simulate the parton showers.
 
All simulated background samples are normalised to their cross-sections,
computed to the highest order available in perturbation theory.
The top-quark mass is set to $m_t=172.5~\gev$ in all
simulated samples. The \evtgen v1.6.0 program~\cite{Lange:2001uf} was used to simulate the decay of
bottom and charm hadrons for all event generators except \sherpa.
 
The nominal \ttbar signal and background samples were processed through the ATLAS
simulation software~\cite{Aad:2010ah} based on GEANT4~\cite{GEANT4}.
Some of the alternative \ttbar samples used to evaluate systematic uncertainties were
processed through a fast detector simulation making use of parameterised showers
in the calorimeters~\cite{AFII}.
Corrections are applied to the simulated events so that the selection efficiencies, energy scales and resolutions of particle candidates match those determined from data control samples.
 
 
\section{Object selection}
\label{sec:selection}
The following sections describe the detector- and particle-level objects
used in the inclusive and fiducial cross-section measurements.
 
\subsection{Detector-level objects}
Electron candidates are reconstructed from energy clusters in the EM calorimeter that match a
reconstructed track.
Electrons are identified with a likelihood method~\cite{Aad:2019tso}, and are required to
meet the tight identification criterion based on shower shapes in
the EM calorimeter, track quality and detection of transition radiation
produced in the TRT.
Electrons are required to have a calorimeter cluster satisfying $|\eta_{\textrm{clust}}|<2.47$. Additionally, electrons in the
transition region between barrel and endcap calorimeters with $1.37 < |\eta_{\textrm{clust}}| < 1.52$
are excluded.
The electron candidates have to pass \pt- and $\eta$-dependent isolation requirements
based on the track and calorimeter activity around them.
Muons are reconstructed using information from both the inner detector
and the muon spectrometer. Muon candidates are required to have $|\eta| < 2.5$, to pass
medium quality requirements~\cite{PERF-2015-10} and fulfil isolation criteria based on the calorimeter and
tracking information: the calorimeter cluster energy within a cone of size of $\Delta R=0.2$ around the muon track divided by the
muon~\pt must be smaller than 0.15 and the ratio of the summed
transverse momenta of additional tracks within a cone of $\Delta R=0.3$ to the muon~\pt must be
smaller than 0.04.
Selected electrons (muons) must have a transverse impact parameter
significance $|d_0/\sigma_{d_{0}}| < 5\,(3)$ and a longitudinal impact
parameter $|z_0 \sin\theta| < 0.5$~mm relative to the event's primary vertex~\cite{ATL-PHYS-PUB-2015-026}.
 
Jets are formed from clusters of topologically connected calorimeter cells~\cite{PERF-2014-07}
using the anti-$k_t$
jet algorithm~\cite{Cacciari:2008gp} with the radius parameter $R=0.4$ implemented in FastJet~\cite{Fastjet}, and
are calibrated to particle level as described in Ref.~\cite{PERF-2016-04}.
To suppress jets originating from pile-up collisions, cuts on the Jet
Vertex Tagger (JVT)~\cite{PERF-2014-03}
discriminant are applied for jets with~\pt below $120~\gev$.
Jets containing $b$-hadrons are identified ($b$-tagged) via a multivariate algorithm, MV2c10, combining
observables sensitive to lifetimes, production mechanisms, and decay properties of
$b$-hadrons~\cite{Aad:2019aic}.
A working point with an average efficiency of 60\% for $b$-quark-initiated jets in \ttbar events
and rejection factors against light-quark/gluon-initiated jets and $c$-quark-initiated jets of 1200 and 55, respectively, is
used~\cite{PERF-2016-05,ATLAS-CONF-2018-001,ATLAS-CONF-2018-006}.
 
The missing transverse momentum with magnitude, $\met$, is defined as the
negative vector sum of the transverse momenta of the reconstructed and
calibrated physics objects (electrons, photons, hadronically decaying
$\tau$-leptons, jets and muons) and a soft term built from all tracks
that are associated with the primary vertex, but not with these
objects, is included~\cite{PERF-2016-07,ATLAS-CONF-2018-023}.
 
\subsection{Particle-level objects}
\label{sec:ParticleLevelDefinition}
 
Particle-level objects are defined in simulated events by using only stable particles,
i.e. particles with a mean lifetime greater than 30 ps. The fiducial phase space
used for the $\sigma_{\ttbar}$ measurement is defined using a set of requirements
applied to particle-level objects analogous to those used in the selection of
the detector-level objects.
 
Leptons are defined as electrons or muons originating from
$W$ decays, including those from intermediate $\tau$-leptons.
The four-momentum of each charged lepton is summed with the four-momenta
of all radiated photons within a cone of size $\Delta R = 0.1$ about
its direction, excluding photons from hadron decays, to account for
bremsstrahlung.
Leptons are required to have $\pt > 25~\gev$ and $|\eta| < 2.5$.
Jets are defined using the anti-$k_t$ algorithm with a radius parameter of $R = 0.4$.
All stable particles are considered for jet clustering, except for the electrons, muons,
and photons used in the lepton definitions.
Jets are required to have $\pt > 25~\gev$ and $|\eta| < 2.5$ and are identified as $b$-jets
via ghost matching to weakly decaying $b$-hadrons~\cite{Cacciari:2008gp}.
The fiducial region is defined by requiring exactly one electron or muon, and at least four
jets, one or exactly two of which must be identified as $b$-jets.
 
Possible double-counting of objects reconstructed at detector- or
particle-levels satisfying multiple object definitions is resolved using the
same algorithms as in Ref.~\cite{Aad:2019ntk}.
 
\section{Analysis strategy}
\label{sec:event}
\subsection{Event selection}
Selected events are required to have exactly one reconstructed electron
or muon with $\pt > 25~\gev$ for the 2015 data-taking period, $\pt > 27~\gev$
for the 2016 data-taking period and $\pt > 28~\gev$ for the 2017 and 2018
data-taking periods, to account for different single-lepton trigger thresholds.
Events must have at least four reconstructed jets with $\pt > 25~\gev$ and
$|\eta| < 2.5$ with one or exactly two of the reconstructed jets being $b$-tagged.
To suppress the contribution of the multijet background, events in the electron+jets channel are required to have $\met > 30~\gev$ and $m_{\mathrm{T}}(W) > 30~\gev$,
while in the muon+jets channel, due to a smaller contribution of this background, a looser criterion $\met+m_{\mathrm{T}}(W) > 60~\gev$ is applied.\footnote{$m_{\mathrm{T}}(W) = \sqrt{2\pt^{\ell}\met\left(1-\cos{\phi}\right)}$, where $\pt^{\ell}$ is the transverse momentum of the charged lepton and $\phi$ is the opening azimuthal angle between the charged lepton and missing transverse momenta.}
The measurement of the \ttbar cross-section is performed by splitting the selected sample into three
non-overlapping signal regions according to the number of jets and $b$-tagged jets. The
region with the highest background fraction (SR1) is selected by requiring $\ge$ 4 jets and exactly 1 $b$-tagged jet.
The SR2 (SR3) region has exactly 4 ($\ge$ 5) jets, exactly two of which
must be $b$-tagged.
The SR1 and SR2 regions have different sensitivities to the background
and $b$-jet modelling while the SR3 provides information about modelling of
extra radiation in \ttbar events.
 
The number of background events meeting the selection criteria is estimated using MC simulations for all processes
with the exception of a small contribution from multijet events with a non-prompt or
misidentified lepton arising from photon conversions, heavy-flavour hadrons decaying
leptonically, and jets misidentified as leptons.
A data-driven matrix method~\cite{Aad:2019ntk} based on the measurement of lepton selection efficiencies using different identification and isolation criteria is used to estimate this background.
Expected and observed event yields are shown
in Table~\ref{tab:EventYields_prefit} and are in excellent agreement. The expected yields
include all uncertainties described in Section~\ref{sec:systematics}.
 
\begin{table}[!hbpt]
\centering
\begin{small}
\begin{tabular}{l D{,}{\,\pm\,}{-1} D{,}{\,\pm\,}{-1} D{,}{\,\pm\,}{-1}}
\toprule
\empty & \multicolumn{1}{c}{SR1} & \multicolumn{1}{c}{SR2} & \multicolumn{1}{c}{SR3} \\
\midrule
\ttbar & 3\,630\,000,210\,000 & 990\,000,90\,000 & 980\,000,100\,000\\
$W$+jets & 350\,000,160\,000 & 24\,000,10\,000 & 17\,000,\phantom{00}9000 \\
Single top & 255\,000,\phantom{0}31\,000 & 52\,000,\phantom{0}7000 & 37\,000,\phantom{00}8000\\
$Z$+jets $\&$ diboson & 80\,000,\phantom{0}40\,000 & 8000,\phantom{0}4000 & 5800,\phantom{00}3000 \\
$\ttbar X$ & 15\,600,\phantom{00}2100 & 2110,\phantom{00\,}290 & 7200,\phantom{00}1000 \\
Multijet & 210\,000,\phantom{0}80\,000 & 28\,000,10\,000 & 22\,000,\phantom{00}8000\\
\midrule
Total prediction & 4\,540\,000,310\,000 & 1\,110\,000,100\,000 & 1\,070\,000,100\,000\\
Data & \multicolumn{1}{c}{4\,540\,886} & \multicolumn{1}{c}{1\,100\,558} & \multicolumn{1}{c}{1\,103\,317} \\
\bottomrule
\end{tabular}
\caption{Expected event yields including all uncertainties after the event selection compared to data
in the three signal regions. The $\ttbar X$ category contains $\ttbar V$
and $\ttbar H$ contributions. }
\label{tab:EventYields_prefit}
\end{small}
\end{table}
 
\subsection{Observables used in the fit}
\label{sec:variables}
 
The \ttbar cross-section is extracted from a simultaneous profile-likelihood fit of data distributions to the sum of signal and background distributions in the three regions.
Each region exploits a different fit variable.
In SR1, the aplanarity ($A$) is used, as was done in previous \ttbar cross-section measurements~\cite{Aaboud:2017cgs,Abazov:2007kg}. It is defined entirely with jet information as $A = \frac{3}{2}\lambda_3$,
where $\lambda_3$ is the smallest eigenvalue of the sphericity
tensor, $S^{\alpha\beta}$ \cite{Tensor1,Tensor2}.\footnote{The $S^{\alpha\beta} = \frac{\sum_i p_i^{\alpha}p_i^{\beta}}{\sum_i |p_i|^2}$, where $p_i$ represents the three-momentum of
jet $i$; $\alpha, \beta \in x, y, z$ and the sum runs over all jets.}
In SR2, the minimum lepton--jet mass, $m_{\ell j}^{\text{min}}$, calculated as the minimum
invariant mass over all lepton--jet pairs, is exploited.
In SR3, a system likely originating from a hadronically decaying top quark is constructed.
It consists of a $b$-tagged jet and two other jets, corresponding to the permutation with the highest \pt for the vector sum of four momenta of the three constituent jets.
The average angular distance between the three constituent jets,
$\Delta R_{bjj}^{\text{avg}}$, is computed and
used in the fit.
The choice of variables is driven by their ability to separate \ttbar signal from the backgrounds,
the reduced sensitivity to jet-related experimental and \ttbar modelling uncertainties
achieved by exploiting ratios of jet momenta ($A$) or angular information
($\Delta R_{bjj}^{\text{avg}}$), and good agreement between the prediction and data.
There is no single variable that satisfies these requirements in all three regions.
 
 
\section{Systematic uncertainties}
\label{sec:systematics}

Several sources of systematic uncertainties affect the fiducial and
inclusive \ttbar cross-section
measurements by changing the estimated signal and background
rates and the shapes of the distributions used in the fit.
All uncertainties are treated as correlated
between signal regions, unless explicitly specified otherwise.
They can be classified into experimental and
modelling uncertainties in the \ttbar signal and in backgrounds.
 
\subsection{Experimental uncertainties}
The uncertainty in the combined 2015--2018 integrated luminosity
(${\mathcal{L}_{\text{int}}}$) is 1.7\%~\cite{ATLAS-CONF-2019-021},
obtained using the LUCID-2 detector~\cite{LUCID2} for the primary luminosity measurements.
 
Reconstruction, identification, isolation and trigger performance for electrons
and muons differ
between data and MC simulations. Scale factors are applied to simulated events to
correct for the differences.
These scale factors, as well as the lepton momentum scale and resolution,
are assessed using $Z\rightarrow \ell^+\ell^-$ events in
simulation and data with methods similar to those described in
Refs.~\cite{Aad:2019tso,PERF-2015-10}. The associated systematic uncertainties
are propagated to the distributions used in the fit. Their combined effects
on the cross-section measurement are referred to as ``Muon reconstruction'' and
``Electron reconstruction'' in Table~\ref{tab:sys_impact}.
 
The jet energy scale (JES) is calibrated using a combination of test beam data, simulation
and \emph{in situ} techniques~\cite{PERF-2016-04}.
Its uncertainty is decomposed into a set of 29 uncorrelated components, with
contributions from pile-up, jet flavour composition, single-particle response,
and effects of jets not contained within the calorimeter.
The uncertainty of the jet energy resolution (JER) is represented by
eight components accounting for jet-$\pt$ and $\eta$-dependent
differences between simulation and data ~\cite{Aad:2011he}.
The uncertainty in the efficiency to pass the JVT requirement for pile-up
suppression is also considered~\cite{PERF-2014-03}. The combined effect
on the cross-section measurement of jet-related uncertainties is referred
to as ``Jet reconstruction'' in Table~\ref{tab:sys_impact}.
 
The uncertainties in the $b$-tagging calibration are determined separately for $b$-jets,
$c$-jets and light-flavour-jets~\cite{Aad:2019aic, ATLAS-CONF-2018-001,ATLAS-CONF-2018-006}
using an 85-component breakdown (45 for $b$-jets, 20 for $c$-jets and 20 for light-flavour jets).
They depend on \pt for $b$- and $c$-jets, and on \pt and $\eta$ for light-flavour jets,
and they account for differences between data and simulation.
The impact of these uncertainties on the cross-section measurement is referred
to as ``Flavour tagging'' in Table~\ref{tab:sys_impact}.
 
The uncertainty in \met due to a possible miscalibration of its soft-track
component is derived from data--simulation comparisons of the \pT balance
between the hard and the soft \met components~\cite{PERF-2016-07}.
To account for the difference in pile-up distributions between the simulation
and data, the pile-up profile in the simulation is corrected to match the one in data.
The uncertainty associated with the correction factor is applied.
The combined impact of the \met and pile-up uncertainties is referred
to as ``$E_{\text{T}}^{\text{miss}}$ + pile-up'' in Table~\ref{tab:sys_impact}.
 
\subsection{Signal modelling}
 
The uncertainty due to missing higher-order QCD corrections in the ME
computation is estimated by independently varying the
renormalisation ($\mu_{\text{R}}$) and factorisation ($\mu_{\text{F}}$)
scales by factors of 2.0 and 0.5 with respect to the central value.
Additionally, uncertainties in the amounts of initial- and final-state radiation (FSR)
from the parton shower are assessed by, respectively varying the corresponding
parameter of the A14 parton shower tune (Var3c)~\cite{ATL-PHYS-PUB-2014-021} and
by varying by factors of 2.0 and 0.5 the scale $\mu_{\text{R}}^{\text{FSR}}$.
All four variations are taken to be uncorrelated between the signal regions but fully correlated across bins in each region. The combined impact of all
scale uncertainties is referred to as ``\ttbar~scale variations'' in
Table~\ref{tab:sys_impact}.
An uncertainty due to the choice of the $h_\text{damp}$ parameter value is determined by comparing
the nominal \ttbar sample with the one produced with the same settings but with the
$h_\text{damp}$ parameter set to 3~$m_t$ and is symmetrised.
 
The level of agreement between data and prediction for the lepton \pt and the leading jet \pt
improves if the top-quark \pt distribution in the nominal \ttbar simulation is
corrected to match the top-quark \pt calculated at NNLO in QCD with NLO electroweak
corrections~\cite{Czakon:2017wor}.
In this analysis, the full difference between the nominal and the reweighted
simulated \ttbar sample is taken as a systematic uncertainty and symmetrised.
This approach is preferable to applying a correction to the nominal simulation
because for some variables the level of agreement between data and prediction deteriorates after applying
the correction.
To avoid double counting, modelling uncertainties, which are evaluated using
alternative samples, are derived as the difference between the nominal and
alternative samples, both reweighted to the top-quark \pt theory prediction.
 
Uncertainties due to the choice of parton shower and hadronisation model
are estimated by comparing the nominal sample from \POWHEGBOX interfaced to \PYTHIA
with an alternative sample generated with the same \POWHEGBOX set-up but
interfaced to \HERWIG~7.0.4~\cite{Bellm:2015jjp,Bahr:2008pv}
with angle-ordered parton shower model, the H7UE tune~\cite{Bellm:2015jjp} and the
MMHT2014LO PDF set~\cite{Harland-Lang:2014zoa}.
Further details about the sample settings can be found in Ref.~\cite{ATL-PHYS-PUB-2018-009}.
The difference between the two models is split into three components.
The first component represents the total \ttbar acceptance in the three
regions (``Shower model incl. acceptance'' in Figure~\ref{fig:Ranking_Incl}).
The second component is sensitive to the \ttbar yield difference in the individual signal regions (``Shower migration parameter'' in Figure~\ref{fig:Ranking_Incl}).
The last component is responsible for the shape effect on the fitted
distributions. It is represented by three nuisance parameters (NPs),
one per region (referred to as ``Shower model shape'' followed by a region
name in Figure~\ref{fig:Ranking_Incl}), to ensure that shape effects are
uncorrelated between the regions since different variables are used in
the fit. All three components are symmetrised. The combined impact of all
uncertainties due to the choice of parton shower and hadronisation model
is referred to as ``\ttbar shower/hadronisation'' in Table~\ref{tab:sys_impact}.
 
The PDF4LHC15 meta-PDFs are used to estimate the systematic effects,
including impact on the acceptance, due to uncertainties in the PDF,
following the updated PDF4LHC15 prescription~\cite{Butterworth:2015oua}. A set
of 30 Hessian eigenvectors corresponding to independent PDF variations is
included in the fit.
The central values of the NNPDF3.0 PDF used to simulate the nominal \ttbar sample
and the PDF4LHC15 set are found to be consistent.
 
\subsection{Background modelling}
 
Uncertainties in the multijet background estimation include
a 50\% uncertainty in the normalisation to cover differences
between the data and the matrix method prediction in various control regions
enriched in multijet background events~\cite{Aad:2019ntk}
and an uncertainty from the choice of parameterisation of the
efficiencies for real and misidentified leptons.
These uncertainties are treated as uncorrelated between all regions and between
electron+jets and muon+jets events due to different composition of the multijet
background in these regions and different
choice of efficiency parameterisation in the electron+jets and muon+jets
channels. The impact of the multijet background estimation uncertainty on the
measurement is referred to as ``Multijet background'' in Table~\ref{tab:sys_impact}.
 
The $tW$ contribution is the largest among the three single-top-quark production channels.
A normalisation uncertainty of 5.4\% is applied to the single-top-quark background,
corresponding to the theoretical uncertainty of the $tW$ cross-section~\cite{Kidonakis:2005kz}.
Similarly to the \ttbar modelling uncertainties, the effects of the $\mu_{\text{R}}$ and $\mu_{\text{F}}$
variations in the ME, the variations of parameters related to initial-
and final-state radiation in the parton shower and the impact of the parton shower choice
are evaluated for the single-top-quark background.
An additional uncertainty arising from the method used to handle interference between
$tW$ and \ttbar\ production is determined by comparing the $tW$
simulated sample that uses the diagram-subtraction method~\cite{Re:2010bp} with the
nominal one based on the diagram-removal technique.
 
Several uncertainties affect the modelling of the $W$+jets background.
Variations of $\mu_{\text{R}}$ and $\mu_{\text{F}}$ are used to derive
the $W$+jets normalisation uncertainties in each region. They amount to
about 45\% and are treated as uncorrelated between the regions
selected with 1-$b$-tag (SR1) and 2-$b$-tag (SR2 and SR3) requirements.
The effects on the shape of the distributions arising from
the $\mu_{\text{R}}$ and $\mu_{\text{F}}$ variations, from
the choice of ME to parton-shower CKKW matching
scale~\cite{Catani:2001cc, Krauss:2002up} and from the scale used for the resummation
of soft-gluon emission in the nominal sample are also included.
 
A normalisation uncertainty of 50\% is applied to the combined $Z$+jets and diboson background
based on the studies of the $\mu_{\text{R}}$ and $\mu_{\text{F}}$ variations for the
$W$+jets process. A normalisation uncertainty of 13.3\% is applied~\cite{deFlorian:2016spz} to the
$\ttbar X$ contribution, based on the theoretical cross-section uncertainties for the
$\ttbar V$ and $\ttbar H$ processes.
 
For the backgrounds, the systematic uncertainties due to the PDF choice are found to
be negligible. The combined effect on the measured cross-section of all MC simulation background modelling uncertainties is referred to as ``MC background modelling'' in Table~\ref{tab:sys_impact}.
 

\section{Extraction of the \ttbar cross-section}
\label{sec:extraction}
 
Events fulfilling the criteria described in Section~\ref{sec:event} are used to perform measurements of the fiducial and inclusive
\ttbar cross-sections from a profile-likelihood fit to data.
The fit uses the distributions of variables described in Section~\ref{sec:variables} in three signal regions, and the systematic uncertainties (see Section~\ref{sec:systematics}) are included in the fit as NPs.
Statistical uncertainties in each bin due to the limited size of the simulated samples
are taken into account by dedicated nuisance parameters using the Barlow-Beeston
technique~\cite{Barlow:249779} and their effect on the measurement is referred to as
``Simulation stat. uncertainty'' in Table~\ref{tab:sys_impact}.
 
The cross-section for producing \ttbar events in the fiducial region,
$\sigma_{\text{fid}}$, is defined as
$\sigma_{\text{fid}} = {\nu_{\text{fid}}}/{\mathcal{L}_{\text{int}}}$,
where $\nu_{\text{fid}}$ is the number of \ttbar events in the fiducial
volume determined by the fit. The inclusive cross-section, $\sigma_{\text{inc}}$,
is related to the fiducial one via
$\sigma_{\text{fid}} = A_{\text{fid}} \times \sigma_{\text{inc}}$,
where $A_{\text{fid}} = {N_{\text{fid}}}/{N_{\text{tot}}}$ is
the fiducial acceptance with $N_{\text{fid}}$ ($N_{\text{tot}}$)
being the number of \ttbar events obtained from a simulated signal sample
after (before) applying the particle-level selection.
For the $\sigma_{\text{fid}}$ measurement, all samples of simulated events used to
evaluate the \ttbar modelling uncertainties are scaled to the same fiducial
acceptance, defined in Section~\ref{sec:ParticleLevelDefinition}.
The fiducial acceptance is evaluated using the nominal \ttbar sample reweighted to match the top-quark~\pt theoretical calculation to be consistent with the treatment of the alternative \ttbar samples. 
Such scaling ensures that in each signal region the remaining normalisation uncertainties
from \ttbar modelling correspond to the uncertainties in the correction factor
$C = {N_{\text{reco}}}/{N_{\text{fid}}}$, where $N_{\text{reco}}$ is the number of selected
events in a given region. The scaled distributions enter the fit to
measure $\sigma_{\text{fid}}$,
thus reducing the impact of \ttbar modelling uncertainties by
reducing the normalisation effects. For the $\sigma_{\text{inc}}$
extraction, the \ttbar modelling uncertainties include the uncertainties
corresponding to the extrapolation of each systematic uncertainty component to the full
phase space.
The acceptance $A_{\text{fid}}$ for different systematic variations of the \ttbar model is shown in Table~\ref{tab:results_acceptance_part}. The PDF uncertainty
is calculated following the PDF4LHC15 prescription as a
sum in quadrature of uncertainties from 30 independent PDF variations.
The relative acceptance uncertainty in the propagation of the fiducial cross-section to the full phase space for the nominal \ttbar model is $^{+1.9}_{-2.2}\%$. 
 
\begin{table}[!hbpt]
\centering
\begin{tabular}{lll}
\toprule
\textbf{Generator set-up}       & $\boldsymbol{A_{\text{fid}}}$ \textbf{[\%]}  & $\boldsymbol{\frac{A^{\text{alt}}_{\text{fid}}-A^{\text{nom}}_{\text{fid}}}{A^{\text{nom}}_{\text{fid}}}}$ \textbf{[\%]} \\
\midrule
\textsc{Powheg+Pythia} nominal     & 13.50 & $\phantom{-}0.00$             \\
\textsc{Powheg+Pythia} top-quark \pt reweighted & 13.40 & $-0.75$            \\
\midrule
$\mu_{\text{R}}^{\text{FSR}} \times 2$      & 13.58 & $\phantom{-}1.29$        \\
$\mu_{\text{R}}^{\text{FSR}} \times 0.5$    & 13.18 & $-1.64$           \\
$\mu_{\text{R}} \times 2$       & 13.37 & $-0.25$            \\
$\mu_{\text{R}} \times 0.5$     & 13.45 & $\phantom{-}0.38$           \\
$\mu_{\text{F}} \times 2$       & 13.38 & $-0.15$            \\
$\mu_{\text{F}} \times 0.5$     & 13.43 & $\phantom{-}0.17$       \\
Var3cUp       & 13.46 & $\phantom{-}0.41$         \\
Var3cDown     & 13.35 & $-0.38$                  \\
$h_{\text{damp}} \times 2$       & 13.57 &  $\phantom{-}1.21$       \\
\textsc{Powheg+Herwig}          & 13.44 & $\phantom{-}0.31$         \\
\midrule
PDF4LHC15 variations     &  & $\phantom{-}0.47$   \\
\midrule
Total  &   & $\phantom{-}^{+1.9}_{-2.2}$ \\
\bottomrule
\end{tabular}
\caption{Fiducial acceptances for different \ttbar models, with the variations relative to the nominal model, after applying the particle-level event selection.
The uncertainty in the acceptance due to each systematic
variation ($A^{\text{alt}}_{\text{fid}}$) is computed
with respect to the acceptance obtained from the nominal \ttbar
sample reweighted to the NNLO theory prediction of the top-quark \pt
given in the second row ($A^{\text{nom}}_{\text{fid}}$). The PDF uncertainty is a sum in quadrature of uncertainties from 30
independent PDF variations in the PDF4LHC15 prescription.
The last row shows the total relative uncertainty in the nominal acceptance.}
\label{tab:results_acceptance_part}
\end{table}

\clearpage

 
\section{Results}
\label{sec:result}
\begin{figure}[!hbpt]
\subfigure[][]{\includegraphics[width=0.32\textwidth]{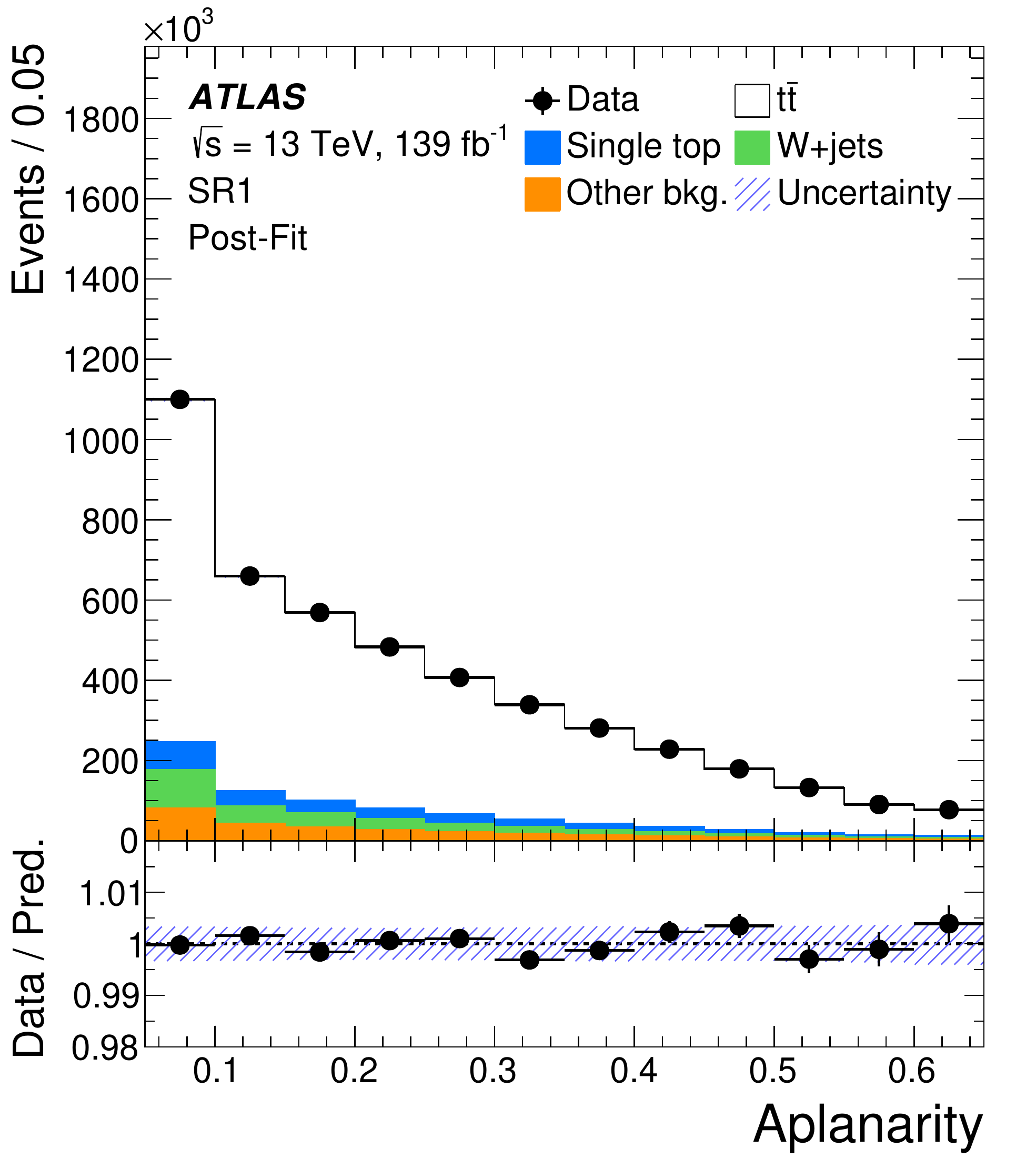}} \hfill
\subfigure[][]{\includegraphics[width=0.32\textwidth]{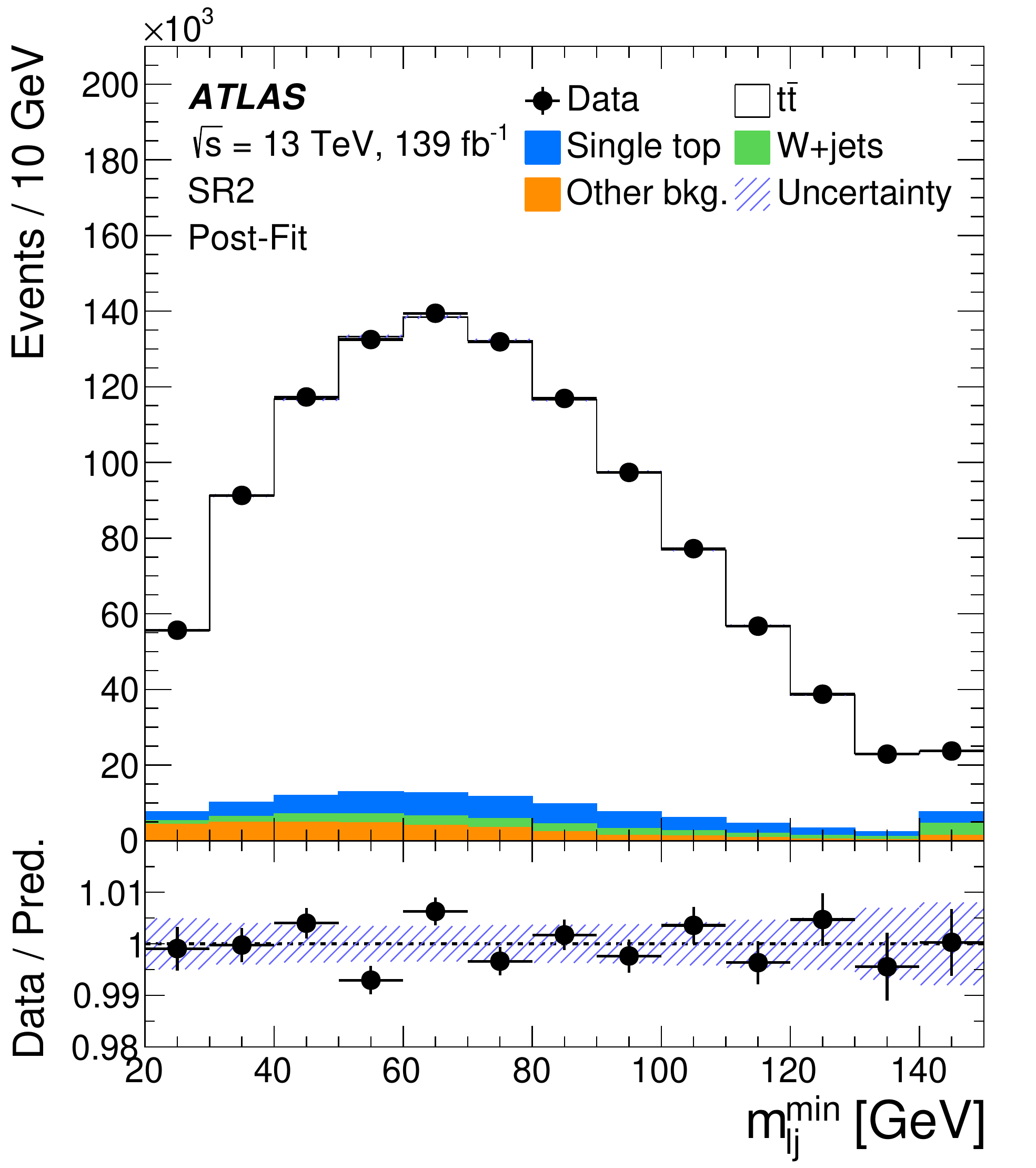}}\hfill
\subfigure[][]{\includegraphics[width=0.32\textwidth]{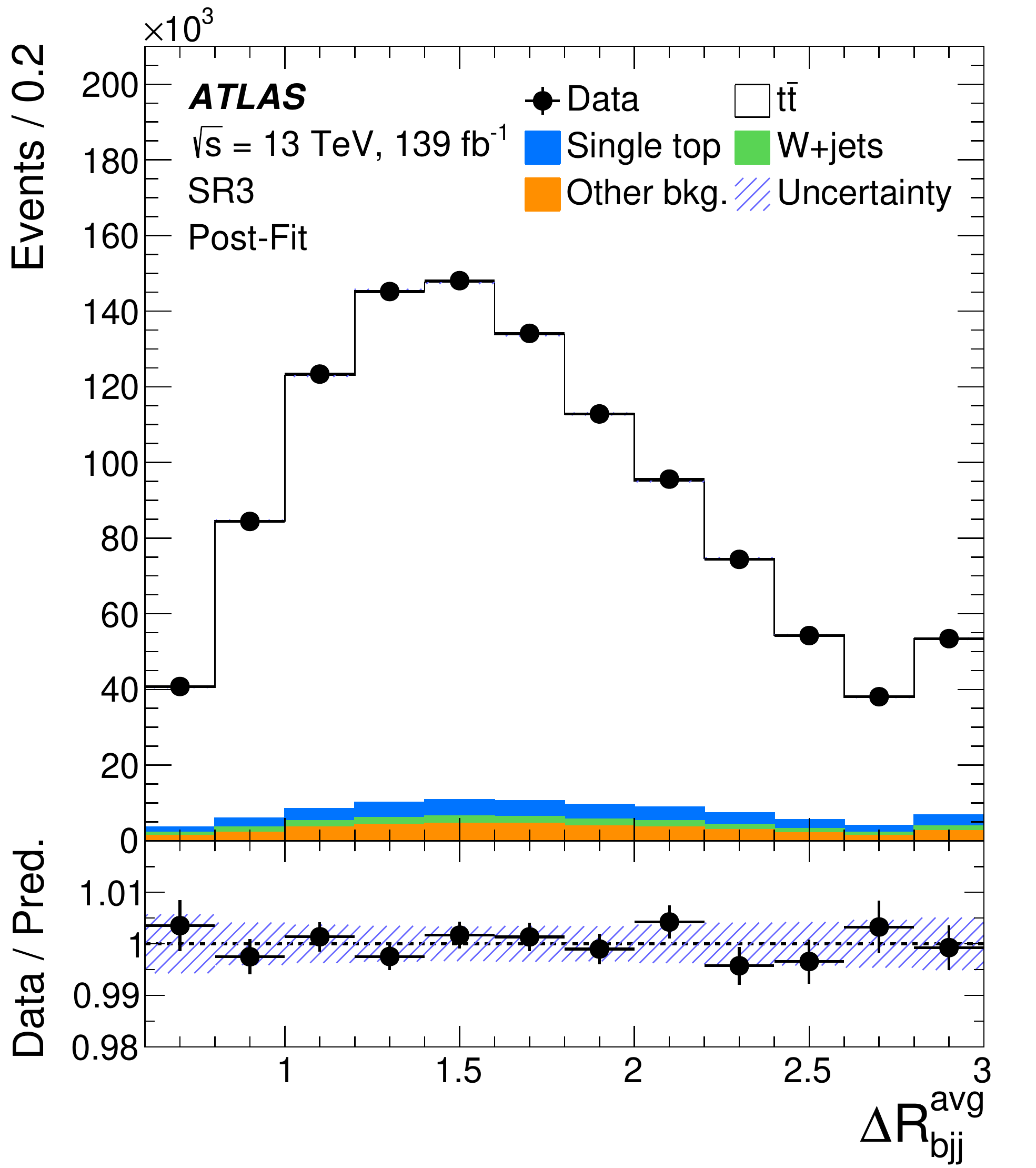}}
\caption{Post-fit distributions of \ttbar signal and backgrounds compared with data for the
observables used in the fiducial cross-section fit. The hatched
bands represent combined statistical and systematic uncertainties, after propagating
the constraints
and correlations obtained from the fit to data. All background categories except single
top and $W$+jets are combined in one category called Other bkg.
The first and last bins contain underflow and overflow
events, respectively.}
\label{fig:post_fit_fid}
\end{figure}

The \ttbar fiducial cross-section is found to be
\begin{eqnarray*}
\sigma_{\text{fid}} = 110.7 \pm 0.05~\text{(stat.)}~^{+4.5}_{-4.3}~\text{(syst.)} \pm
1.9~\text{(lumi.)}~\text{pb} = 110.7\pm 4.8~\text{pb}.
\end{eqnarray*}
Here, the luminosity uncertainty is obtained by repeating the fit, fixing the corresponding
nuisance parameter, and subtracting in quadrature the resulting uncertainty from the total
uncertainty of the nominal fit. The systematic uncertainty is determined by subtracting
in quadrature the statistical uncertainty, obtained from a fit where
all NPs are fixed to the values determined by the fit (post-fit), and the luminosity
uncertainty, from the total uncertainty.
Figure~\ref{fig:post_fit_fid} displays the post-fit distributions of the observables used in
the fit in each region.
 
\begin{figure}[!hbpt]
\subfigure[][]{\includegraphics[width=0.32\textwidth]{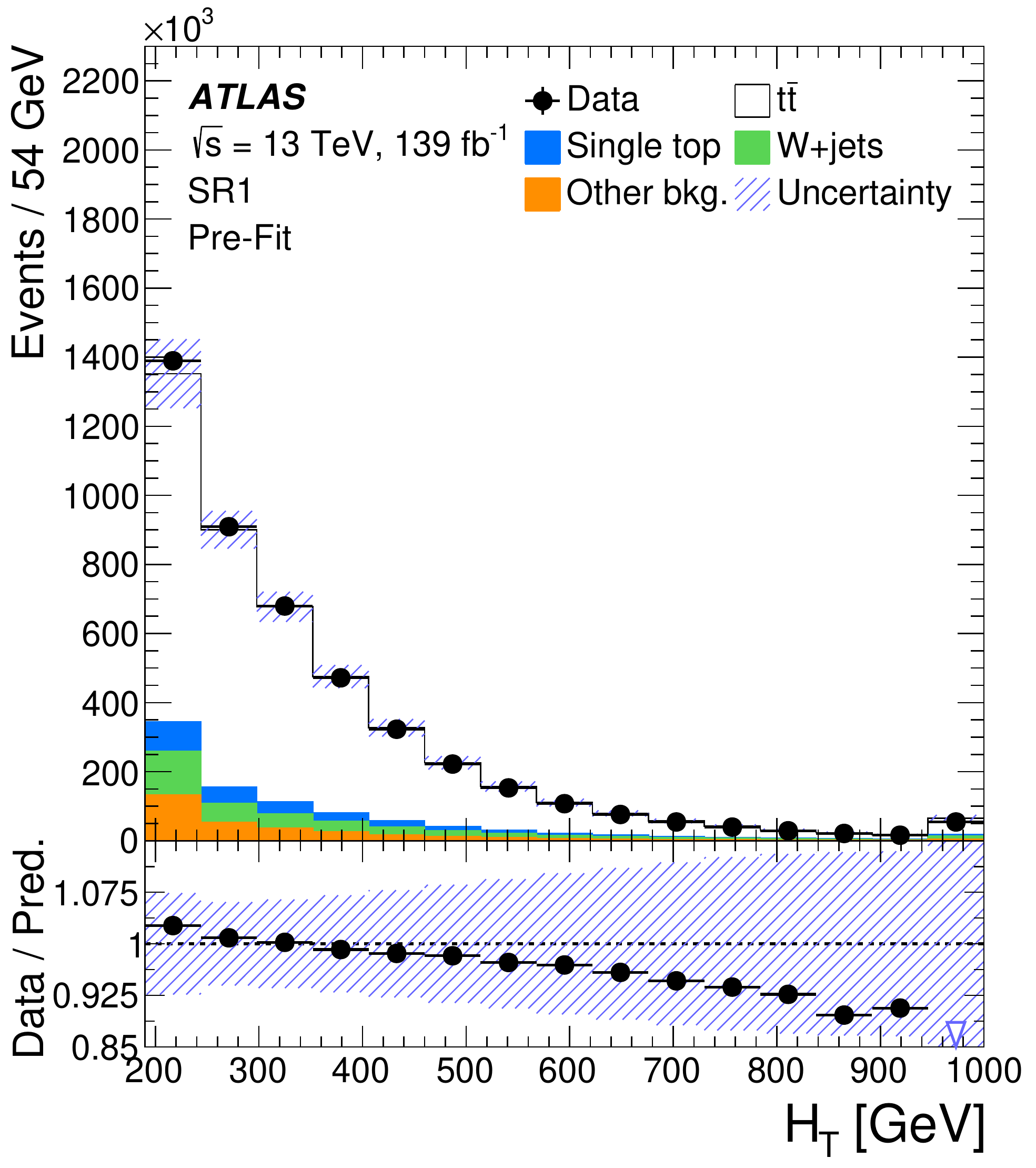}} \hfill
\subfigure[][]{\includegraphics[width=0.32\textwidth]{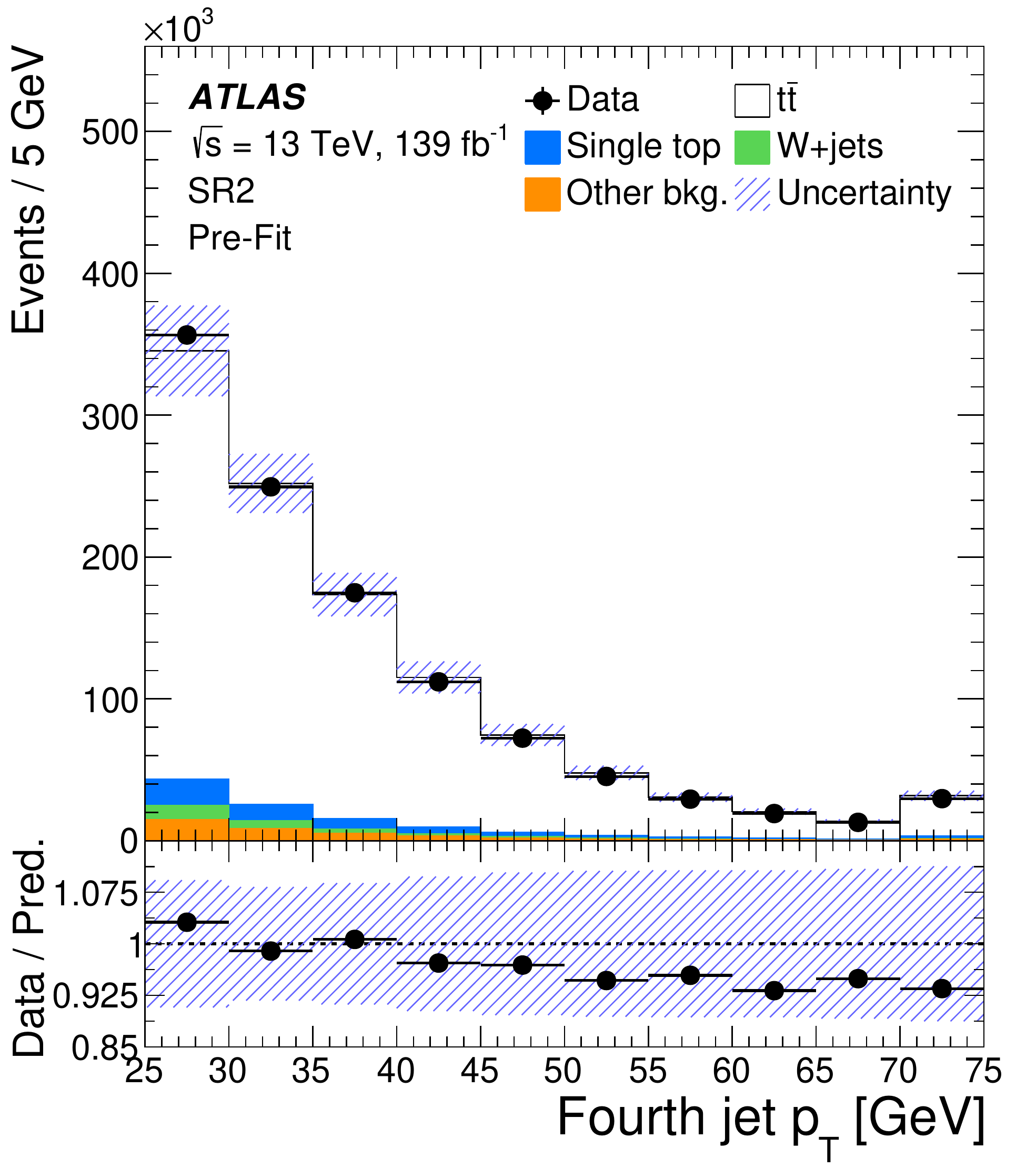}}\hfill
\subfigure[][]{\includegraphics[width=0.32\textwidth]{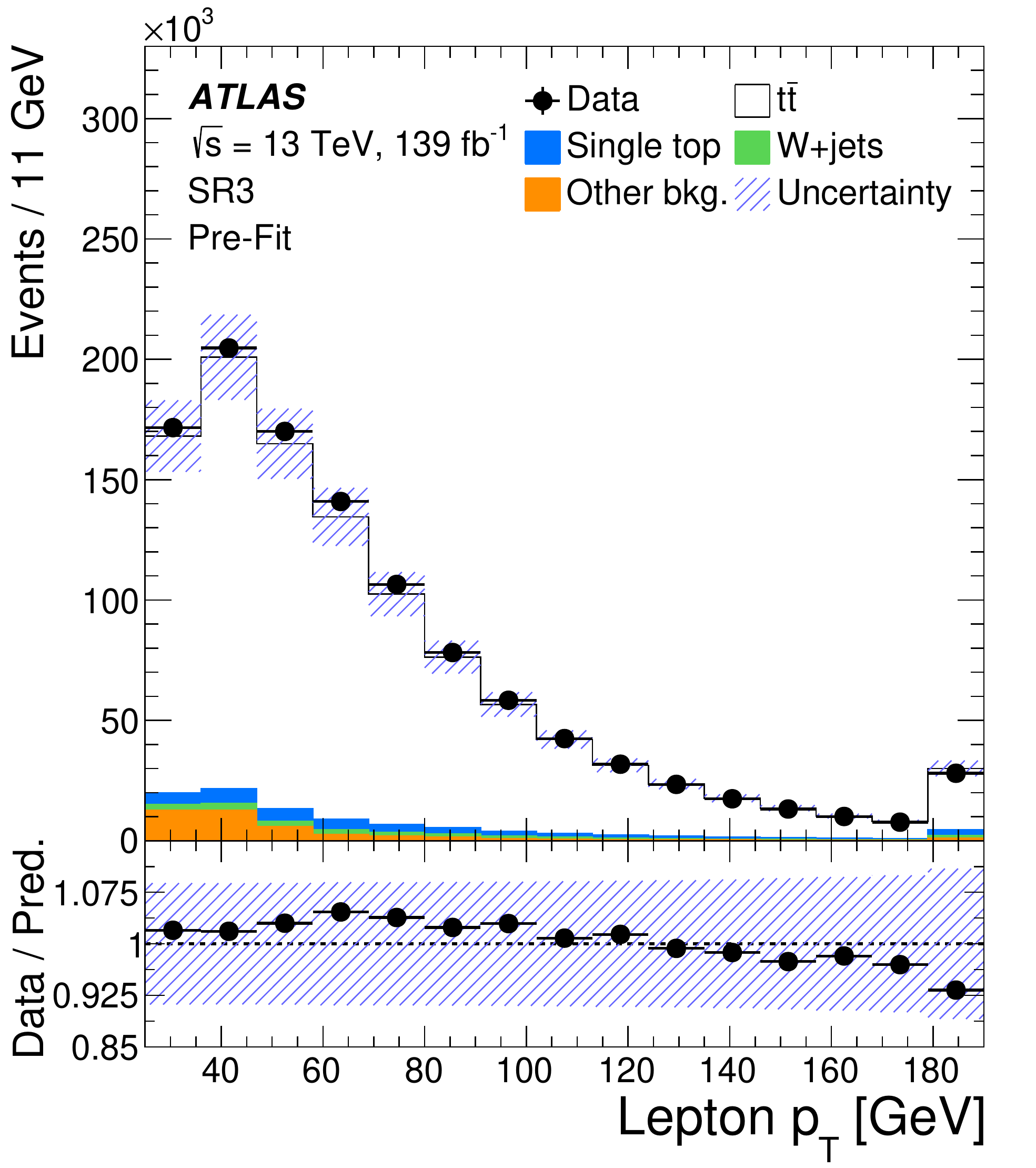}}\\
\subfigure[][]{\includegraphics[width=0.32\textwidth]{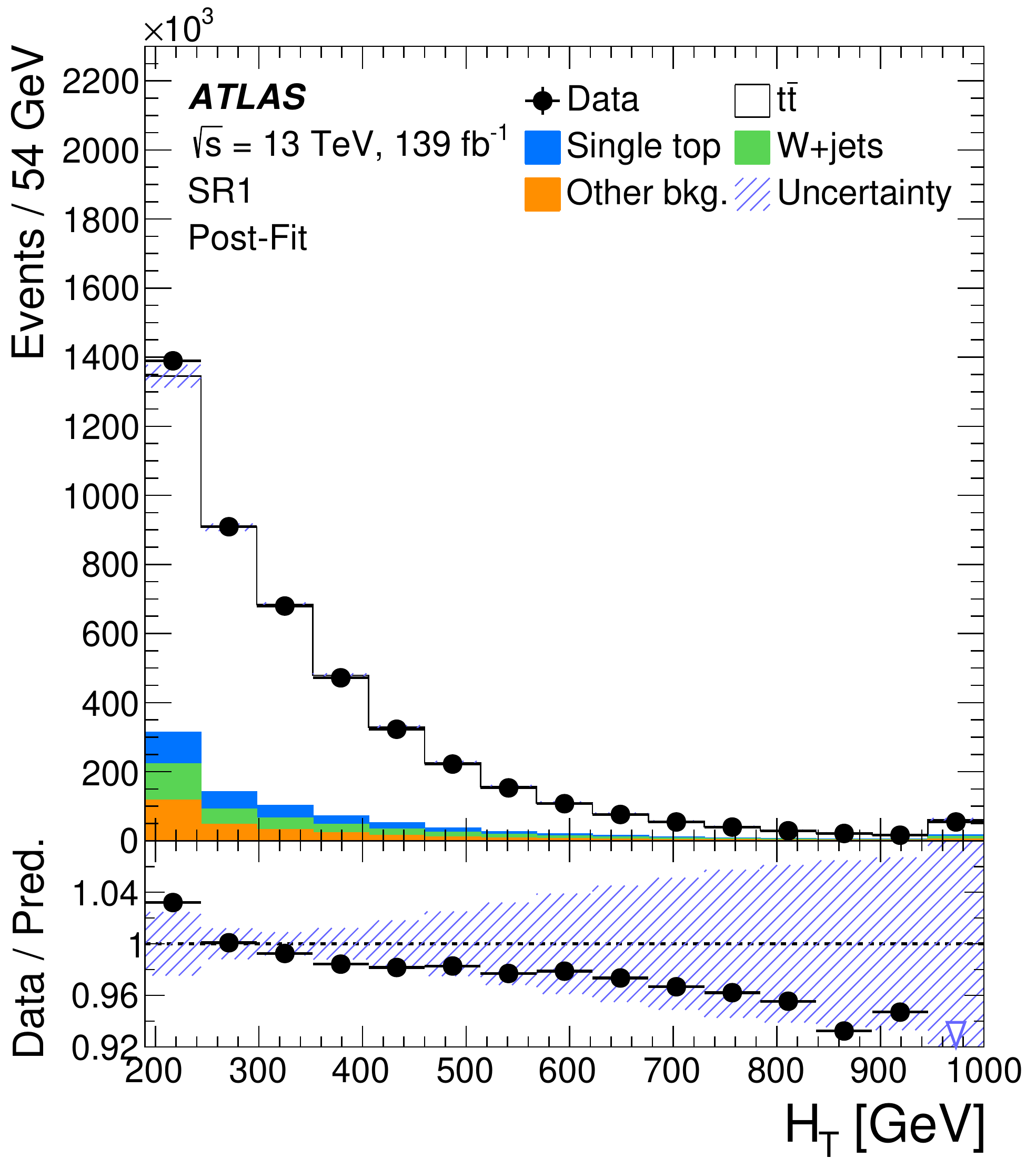}} \hfill
\subfigure[][]{\includegraphics[width=0.32\textwidth]{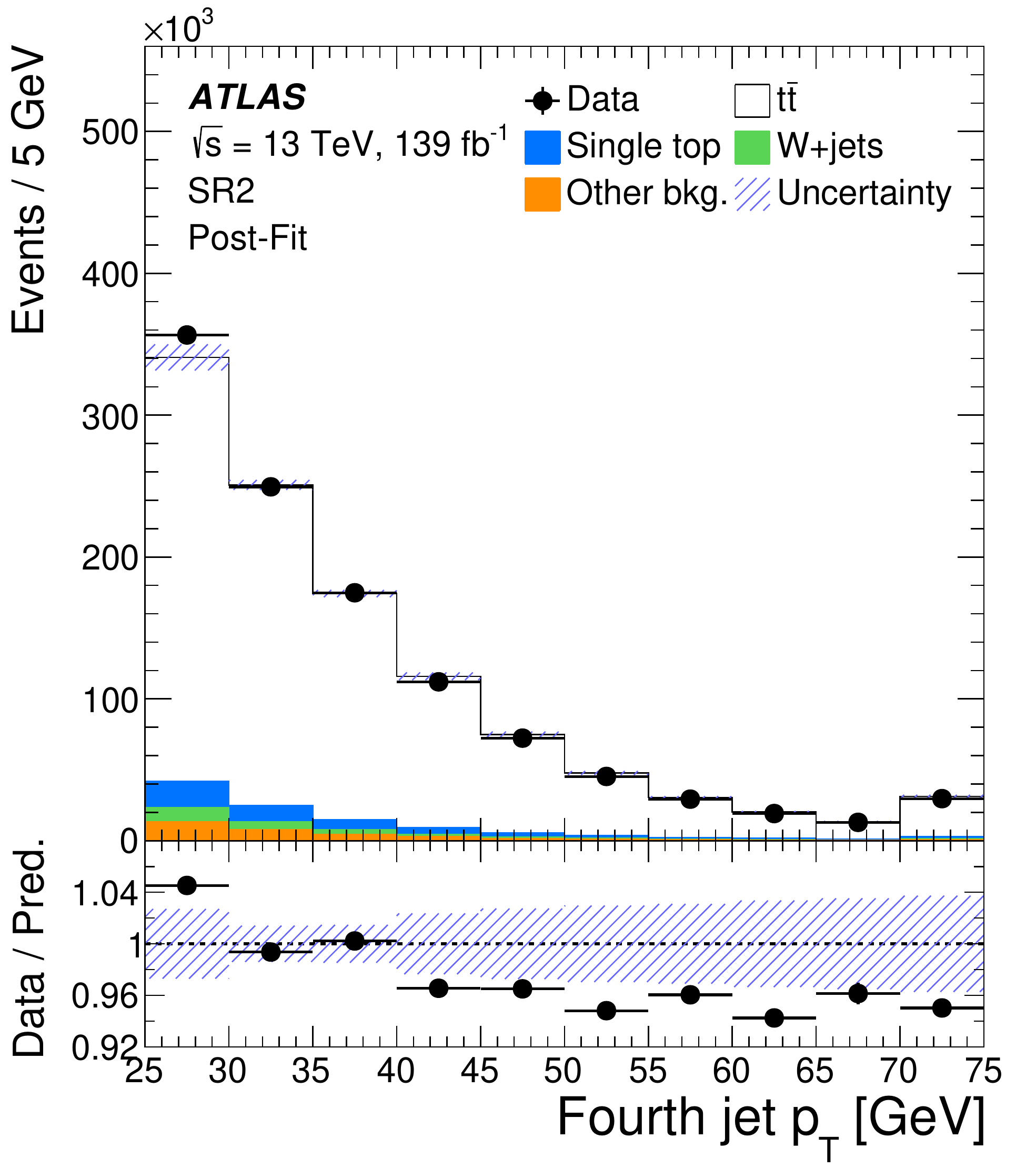}}\hfill
\subfigure[][]{\includegraphics[width=0.32\textwidth]{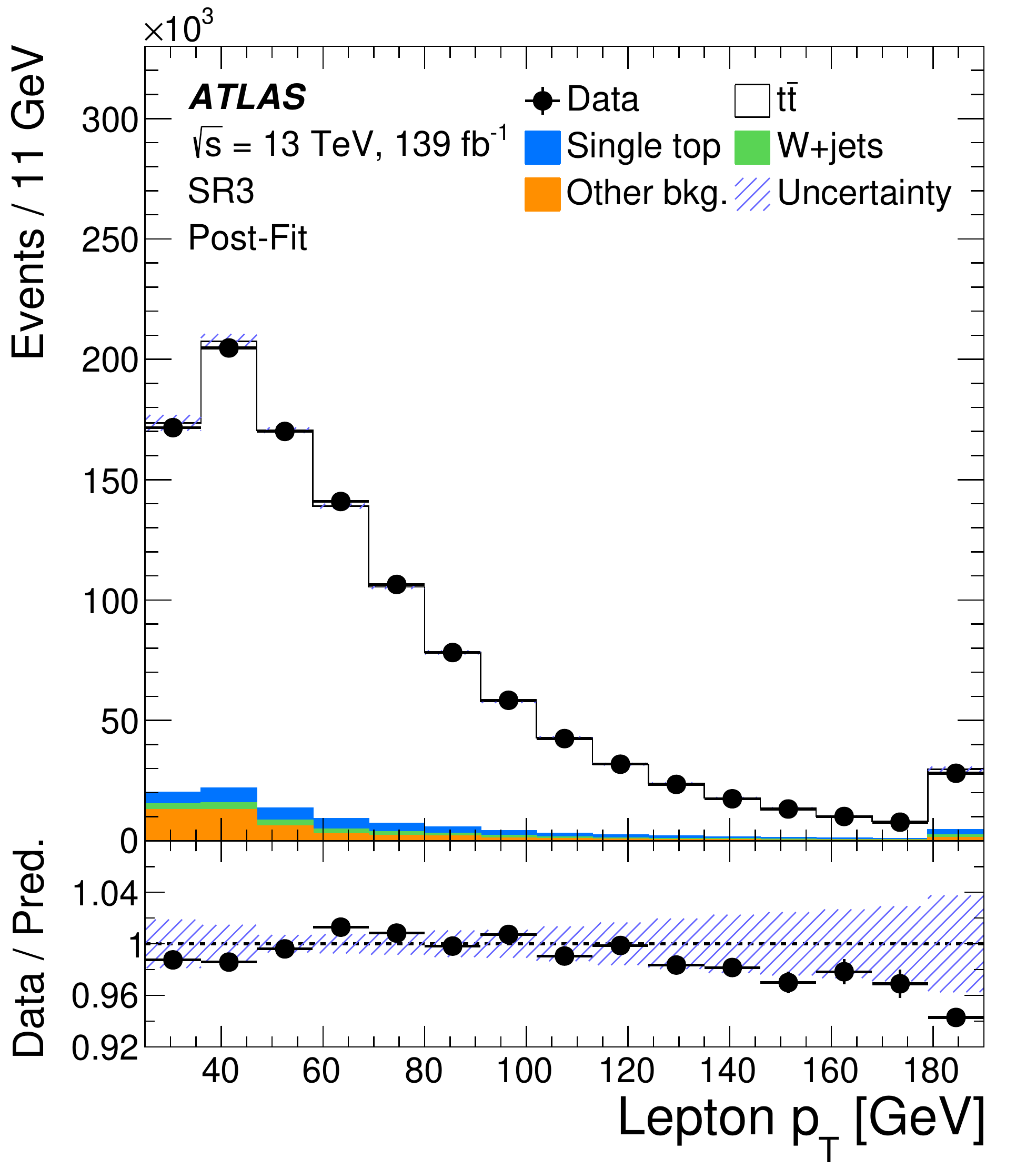}}
 
\caption{Pre-fit (top) and post-fit (bottom) distributions of the scalar sum of jet transverse
momenta in the event ($H_{\textrm{T}}$) in SR1 (left), the fourth largest jet \pt in SR2 (middle) and
the lepton \pt in SR3 (right) for the fiducial cross-section measurement. The hatched bands represent
combined statistical and systematic
uncertainties. The first and last bins contain underflow and overflow events, respectively.}
\label{fig:post_fit_fid_validation}
\end{figure}

Figure~\ref{fig:post_fit_fid_validation} shows pre- and post-fit distributions of one kinematic variable per region, which
is not included in the fit, demonstrating that the level of agreement between the prediction
and the data improves after the fit.
The $H_\textrm{T}$ distribution shows a difference between prediction and data, which is covered by the uncertainties both before and after the fit.
This feature has no effect on the variables used in the fit or on the result.
The effect of the residual disagreement in the distribution of the fourth largest jet \pt in
SR2, which is not
fully covered by the post-fit uncertainty band, is tested as follows.
Pseudo-data are created by reweighting the detector-level prediction for events passing the selection to match the corresponding distribution in data in SR2, and the \ttbar cross-section is extracted.
No significant impact on the measured cross-section is observed.
 
Using the measured fiducial cross-section and the acceptance with its uncertainty from
Table~\ref{tab:results_acceptance_part}, and assuming that the uncertainties of
the $A_{\text{fid}}$ are not correlated with those obtained in the fit,
the \ttbar cross-section extrapolated to the full phase space is
\begin{eqnarray*}
\sigma_{\text{inc}}^{\text{ext}} = 820  \pm 0.4~\text{(stat.)} \pm 37~\text{(syst.)}
\pm 14~\text{(lumi.)}~\text{pb} = 820\pm40~\text{pb}.
\end{eqnarray*}
 
The \ttbar cross-section in the full phase space, referred to as inclusive cross-section,
measured in the dedicated fit is
\begin{eqnarray*}
\sigma_{\text{inc}} =
830  \pm 0.4~\text{(stat.)} \pm 36~\text{(syst.)} \pm 14~\text{(lumi.)}~\text{pb}
= 830\pm 38~\text{pb}.
\end{eqnarray*}
 
The two results are compatible within the uncertainties
and are in agreement with the theoretical NNLO + NNLL prediction for the
top-quark mass of 172.5~\GeV. The difference between the central values arises
from the different assumptions related to the \ttbar modelling uncertainties.
For the inclusive measurement, the alternative models are assumed to have
the same $\sigma_{\ttbar}$
in the full phase space, while for the fiducial measurement they are assumed to
have the same cross-section after applying the fiducial selection.
This results in different normalisation components of the signal modelling
uncertainties, leading to different impacts of these uncertainties on the measured
cross-section for the same post-fit values of the corresponding nuisance parameters.
 
The dependence of the measured inclusive \ttbar cross-section on $m_t$
is determined by repeating the fit to data after replacing the nominal input \ttbar
distributions by those from the
samples generated with the same set-up as the nominal but with
$m_t = 171, 172, 173$ and 174~\GeV ,
assuming that the \ttbar modelling uncertainties are independent of $m_t$.
The dependence is found to
be $1/{\sigma_{\text{inc}}}\times {\text{d}\sigma_{\text{inc}}} / {\text{d}m_{t}}=-1.7\%/\gev$.
 
\begin{figure}[!hbpt]
\centering
\includegraphics[width=0.7\textwidth]{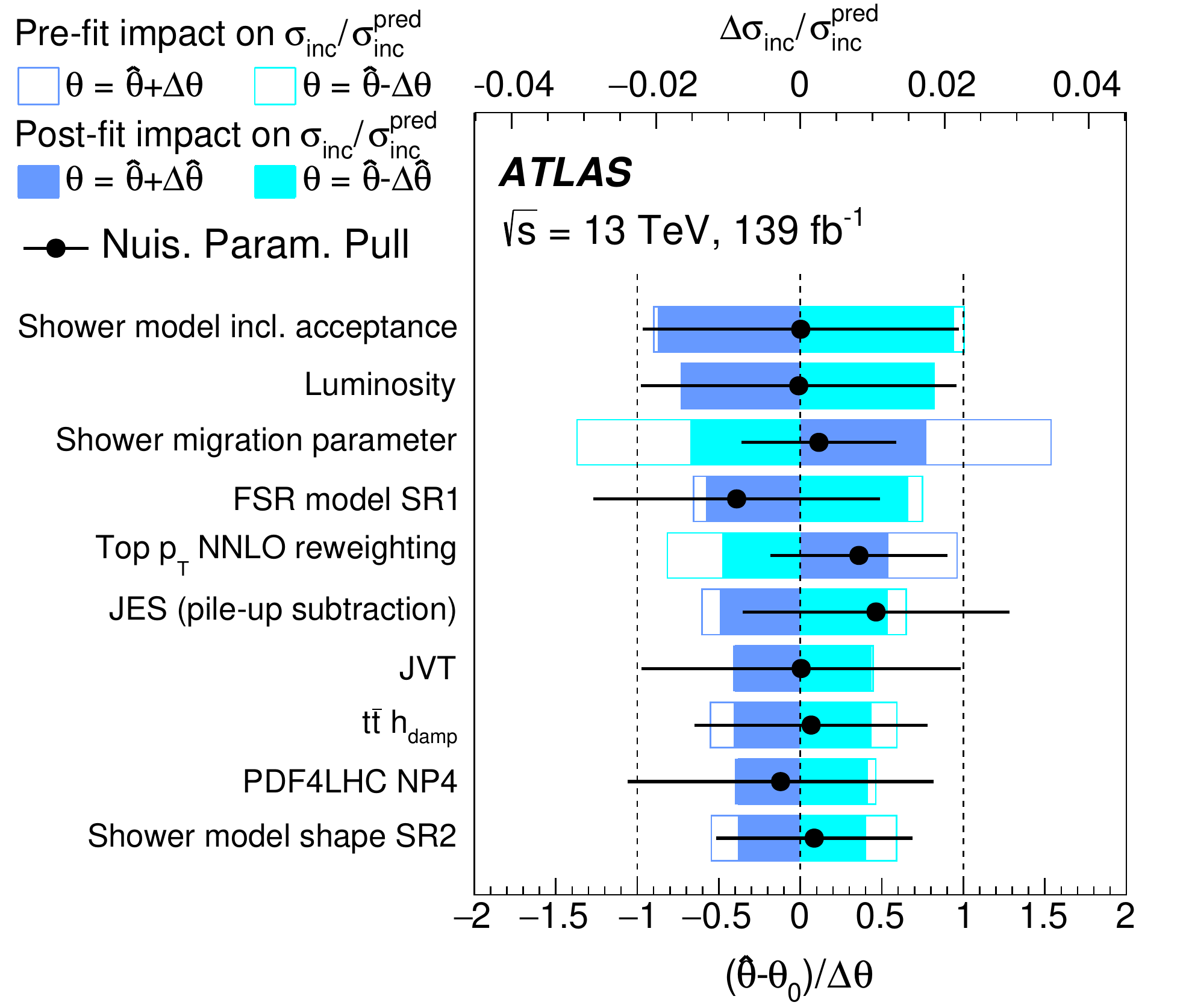}
\caption{Ranking plot showing the effect of the ten most important systematic
uncertainties on the measured cross-section, normalised to the predicted
value, in the inclusive fit to data. The impact of each NP,
$\Delta \sigma_{\text{inc}}/\sigma^{\text{pred.}}_{\text{inc}}$, is computed by
comparing the nominal best-fit value of
$\sigma_{\text{inc}}/\sigma^{\text{pred}}_{\text{inc}}$ with the result of the fit
when fixing the considered nuisance parameter to its best-fit value, $\hat{\theta}$,
shifted by its pre-fit and post-fit uncertainties $\pm \Delta \theta$ ($\pm \Delta \hat{\theta})$. The empty boxes
show the pre-fit impact while the filled boxes show the post-fit impact of each
nuisance parameter on the result. The black dots represent the post-fit value (pull)
of each NP where the pre-fit value is subtracted, while the black line represents
the post-fit uncertainty normalised to the pre-fit uncertainty. The ``JES
(pile-up subtraction)'' is one of the 29 components of the JES uncertainty,
the ``FSR model SR1'' is the FSR scale uncertainty in SR1 and
the ``PDF4LHC NP4'' is one of the 30 independent PDF variations. Other
components are described in Section~\ref{sec:systematics}.}
\label{fig:Ranking_Incl}
\end{figure}

Figure~\ref{fig:Ranking_Incl} presents the ranking of the effects of different systematic
uncertainties on the inclusive measurement. The impact of each NP, $\theta$, is computed by
comparing the nominal best-fit value
of $\sigma_{\text{inc}}$ with the result of the fit when fixing the considered nuisance parameter
to its best-fit value, $\hat{\theta}$, shifted by its pre-fit (post-fit)
uncertainties $\pm \Delta{\theta}$ ($\pm \Delta{\hat{\theta}}$).
The ranking plot shows that the uncertainty in $\sigma_{\text{inc}}$ is
dominated by the difference in the \ttbar inclusive acceptance and the
migration parameter between the nominal and the alternative parton shower
and hadronisation model. The NP corresponding to the migration parameter is
constrained, indicating that the normalisation effects of the alternative
model vary significantly between the three regions. In SR1 (SR3), the
alternative model predicts 1.4\% (2.3\%) larger yield while in SR2 it
predicts 7.1\% smaller yield than in the nominal \ttbar simulation.
These variations are much larger than the data uncertainty and allow the
data to constrain this uncertainty. 
To check that this choice for the parameterisation of the parton shower
systematic uncertainty does not affect the result, an alternative parameterisation
is implemented with three normalisation and three
shape NPs uncorrelated between three signal regions. No change in the central value or
total uncertainty is observed, while the parameters show similar level of
constraints and pulls as in the baseline fit.
Other significant contributions to the uncertainty arise from the modelling
of final-state radiation in SR1 and the top-quark \pt model.
As expected, the latter is pulled towards the NNLO prediction, which is approximated here by a
one-dimensional top-quark \pt reweighting.
The uncertainty in the integrated luminosity is the highest-ranked experimental
uncertainty.
 
A breakdown of the contributions from different categories of systematic
uncertainties is
presented in Table~\ref{tab:sys_impact}. The largest uncertainties, in both the fiducial and
inclusive cross-section measurements, arise from the shower and hadronisation modelling and
the scale variations. The source of the largest experimental uncertainty is the jet reconstruction
category which includes uncertainties from jet identification, calibration, resolution and
the JVT requirement.
 
\begin{table}[!hbpt]
\centering
\footnotesize
\begin{tabular}{lll}
\toprule
\textbf{Category}                   & $\boldsymbol{\frac{\Delta \sigma_{\text{fid}}}{\sigma_{\text{fid}}}}$ \textbf{[\%]} & $\boldsymbol{\frac{\Delta \sigma_{\text{inc}}}{\sigma_{\text{inc}}}}$ \textbf{[\%]} \\
\midrule
\multicolumn{3}{c}{\textbf{Signal modelling}} \\
\midrule
$t\bar{t}$ shower/hadronisation & $\pm 2.8$    & $\pm 2.9$    \\
$t\bar{t}$ scale variations     & $\pm 1.4$    & $\pm 2.0$    \\
Top \pt NNLO reweighting        & $\pm 0.4$    & $\pm 1.1$    \\
$t\bar{t}$ $h_{\text{damp}}$    & $\pm 1.5$    & $\pm 1.4$     \\
$t\bar{t}$ PDF                 & $\pm 1.4$    & $\pm 1.5$     \\
 
\midrule
\multicolumn{3}{c}{\textbf{Background modelling}} \\
\midrule
MC background modelling    & $\pm 1.8$    & $\pm 2.0$     \\
Multijet background        & $\pm 0.8$    & $\pm 0.6$      \\
\midrule
\multicolumn{3}{c}{\textbf{Detector modelling}} \\
\midrule
Jet reconstruction         & $\pm 2.5$     & $\pm 2.6$     \\
Luminosity                 & $\pm 1.7$     & $\pm 1.7$    \\
Flavour tagging            & $\pm 1.2$     & $\pm 1.3$     \\
$E_{\text{T}}^{\text{miss}}$ + pile-up & $\pm 0.3$     & $\pm 0.3$      \\
Muon reconstruction        & $\pm 0.6$     & $\pm 0.5$     \\
Electron reconstruction    & $\pm 0.7$     & $\pm 0.6$    \\
Simulation stat. uncertainty & $\pm 0.6$    & $\pm 0.7$     \\
\midrule
Total systematic uncertainty & $\pm 4.3$     & $\pm 4.6$    \\
Data statistical uncertainty     & $\pm 0.05$    & $\pm 0.05$   \\
\midrule
Total uncertainty          & $\pm 4.3$     & $ \pm4.6$    \\
\bottomrule
\end{tabular}
\caption{Impact of different categories of systematic uncertainties
and data statistics on the fiducial and inclusive measurements.
The quoted values are obtained by repeating the fit, fixing a set of nuisance
parameters of the sources corresponding to the considered category, and subtracting
in quadrature the resulting uncertainty from the total uncertainty of the nominal
fit presented in the last line. The total uncertainty is different from the sum in
quadrature of the different components due to correlations between nuisance parameters
built by the fit. The categories are defined in Section~\ref{sec:systematics}.}
\label{tab:sys_impact}
\end{table}
 
Several tests were performed to check the stability of the result.
To examine the disagreement between data and prediction observed in jet~\pt spectra as illustrated in Figure~\ref{fig:post_fit_fid_validation}, the impact of
changing the minimum jet \pt requirement was studied by repeating the analysis while selecting events
with a minimum jet~\pt of 30~\gev{} and 35~\gev{} instead of 25~\gev.
In both cases, the measured cross-section changed by less than 2\% and did not show a
trend depending on the jet~\pt cut.
 
The approach to performing ME to parton shower matching differs between NLO
generators and, in general, can be a source of uncertainty. However,
it is not straightforward to separate the effect of the algorithmic difference
in the implementation of such matching from other effects when replacing one
ME generator by an alternative one, matched to the same parton shower. This may involve
changes in the parameters of the parton shower that can lead to a much larger
effect than the targeted one. For this reason, the effect of the generator choice
is not included in the fit model. However, its impact on the result
is checked by comparing two alternative \ttbar samples generated with \POWHEGBOX~v2~ and
\MGMCatNLO, both interfaced to \HERWIG~7.1.3~\cite{Reichelt:2017hts}. A symmetrised difference between
these two samples is applied as an additional systematic uncertainty, correlated between
regions. No significant impact on the central value or the uncertainty
is observed for either the inclusive or the fiducial measurements.
 
The stability of the result with respect to the choice of correlation scheme
for the initial- and final-state radiation uncertainties, and for
the $\mu_{\text{R}}$ and $\mu_{\text{F}}$ scale variations, was studied.
In the alternative scheme, the uncertainties were treated as fully correlated across the signal regions.
No effect on either the measured cross-sections or the uncertainties
was observed.
 
\section{Conclusion}
\label{sec:conclusion}
Measurements of the inclusive and fiducial \ttbar production cross-sections are performed in the lepton+jets channel using
proton--proton collision data at $\sqrt{s} = \SI{13}{\TeV}$ recorded by the ATLAS detector at the LHC during 2015--2018,
corresponding to an integrated luminosity of 139~fb$^{-1}$.
The analysis is performed in three regions requiring different jet multiplicities and different numbers of $b$-tagged jets.
The \ttbar production cross-section and its uncertainty are extracted from a profile-likelihood fit to data of the
distributions of discriminating variables in these three regions, assuming $m_t = 172.5~\gev$.
The fiducial cross-section is measured with a precision of 4.3\% to
be $\sigma_{\text{fid}} = 110.7 \pm 4.8~\text{pb} = 110.7 \pm 0.05~\text{(stat.)}{}^{+4.5}_{-4.3}~\text{(syst.)}\pm 1.9~\text{(lumi.)}~\mathrm{pb}$, and
the inclusive cross-section is measured with a precision of 4.6\% to be
$\sigma_{\text{inc}} = 830 \pm 38~\text{pb} =
830 \pm 0.4~\text{(stat.)}\pm 36~\text{(syst.)}\pm 14~\text{(lumi.)}~\mathrm{pb}$.
The inclusive result is in agreement with the theoretical NNLO + NNLL QCD calculation
as well as with the ATLAS measurement in the electron--muon channel
and with CMS measurements.
\section*{Acknowledgements}
 

We thank CERN for the very successful operation of the LHC, as well as the
support staff from our institutions without whom ATLAS could not be
operated efficiently.
 
We acknowledge the support of ANPCyT, Argentina; YerPhI, Armenia; ARC, Australia; BMWFW and FWF, Austria; ANAS, Azerbaijan; SSTC, Belarus; CNPq and FAPESP, Brazil; NSERC, NRC and CFI, Canada; CERN; CONICYT, Chile; CAS, MOST and NSFC, China; COLCIENCIAS, Colombia; MSMT CR, MPO CR and VSC CR, Czech Republic; DNRF and DNSRC, Denmark; IN2P3-CNRS and CEA-DRF/IRFU, France; SRNSFG, Georgia; BMBF, HGF and MPG, Germany; GSRT, Greece; RGC and Hong Kong SAR, China; ISF and Benoziyo Center, Israel; INFN, Italy; MEXT and JSPS, Japan; CNRST, Morocco; NWO, Netherlands; RCN, Norway; MNiSW and NCN, Poland; FCT, Portugal; MNE/IFA, Romania; MES of Russia and NRC KI, Russia Federation; JINR; MESTD, Serbia; MSSR, Slovakia; ARRS and MIZ\v{S}, Slovenia; DST/NRF, South Africa; MINECO, Spain; SRC and Wallenberg Foundation, Sweden; SERI, SNSF and Cantons of Bern and Geneva, Switzerland; MOST, Taiwan; TAEK, Turkey; STFC, United Kingdom; DOE and NSF, United States of America. In addition, individual groups and members have received support from BCKDF, CANARIE, Compute Canada and CRC, Canada; ERC, ERDF, Horizon 2020, Marie Sk{\l}odowska-Curie Actions and COST, European Union; Investissements d'Avenir Labex, Investissements d'Avenir Idex and ANR, France; DFG and AvH Foundation, Germany; Herakleitos, Thales and Aristeia programmes co-financed by EU-ESF and the Greek NSRF, Greece; BSF-NSF and GIF, Israel; CERCA Programme Generalitat de Catalunya and PROMETEO Programme Generalitat Valenciana, Spain; G\"{o}ran Gustafssons Stiftelse, Sweden; The Royal Society and Leverhulme Trust, United Kingdom.
 
The crucial computing support from all WLCG partners is acknowledged gratefully, in particular from CERN, the ATLAS Tier-1 facilities at TRIUMF (Canada), NDGF (Denmark, Norway, Sweden), CC-IN2P3 (France), KIT/GridKA (Germany), INFN-CNAF (Italy), NL-T1 (Netherlands), PIC (Spain), ASGC (Taiwan), RAL (UK) and BNL (USA), the Tier-2 facilities worldwide and large non-WLCG resource providers. Major contributors of computing resources are listed in Ref.~\cite{ATL-SOFT-PUB-2020-001}.
 

\printbibliography
 
\clearpage 
 
\begin{flushleft}
\hypersetup{urlcolor=black}
{\Large The ATLAS Collaboration}

\bigskip

\AtlasOrcid[0000-0002-6665-4934]{G.~Aad}$^\textrm{\scriptsize 102}$,    
\AtlasOrcid[0000-0002-5888-2734]{B.~Abbott}$^\textrm{\scriptsize 128}$,    
\AtlasOrcid{D.C.~Abbott}$^\textrm{\scriptsize 103}$,    
\AtlasOrcid[0000-0002-2788-3822]{A.~Abed~Abud}$^\textrm{\scriptsize 36}$,    
\AtlasOrcid[0000-0002-1002-1652]{K.~Abeling}$^\textrm{\scriptsize 53}$,    
\AtlasOrcid[0000-0002-2987-4006]{D.K.~Abhayasinghe}$^\textrm{\scriptsize 94}$,    
\AtlasOrcid[0000-0002-8496-9294]{S.H.~Abidi}$^\textrm{\scriptsize 166}$,    
\AtlasOrcid[0000-0002-8279-9324]{O.S.~AbouZeid}$^\textrm{\scriptsize 40}$,    
\AtlasOrcid{N.L.~Abraham}$^\textrm{\scriptsize 155}$,    
\AtlasOrcid[0000-0001-5329-6640]{H.~Abramowicz}$^\textrm{\scriptsize 160}$
% ...
\bigskip
\\

% ...
$^{36}$CERN, Geneva; Switzerland.\\
% ...
$^{40}$Niels Bohr Institute, University of Copenhagen, Copenhagen; Denmark.\\
% ...
$^{53}$II. Physikalisches Institut, Georg-August-Universit\"{a}t G\"ottingen, G\"ottingen; Germany.\\
% ...
$^{94}$Department of Physics, Royal Holloway University of London, Egham; United Kingdom.\\
% ...
$^{102}$CPPM, Aix-Marseille Universit\'e, CNRS/IN2P3, Marseille; France.\\
$^{103}$Department of Physics, University of Massachusetts, Amherst MA; United States of America.\\
% ...
$^{128}$Homer L. Dodge Department of Physics and Astronomy, University of Oklahoma, Norman OK; United States of America.\\
% ...
$^{155}$Department of Physics and Astronomy, University of Sussex, Brighton; United Kingdom.\\
% ...
$^{160}$Raymond and Beverly Sackler School of Physics and Astronomy, Tel Aviv University, Tel Aviv; Israel.\\
% ...
$^{166}$Department of Physics, University of Toronto, Toronto ON; Canada.\\
% ...

\end{flushleft}


\end{document}